\documentclass[aps,prb,preprint,groupedaddress]{revtex4}
\usepackage{graphicx}

\begin{document}
\title{Quantum Teleportation in One-Dimensional Quantum Dots System}
\author{Hefeng Wang and  Sabre Kais \footnote{
Corresponding author:  kais@purdue.edu}}
\affiliation{Department of Chemistry, Purdue University, West Lafayette, IN 47907} 
\begin{abstract}
\noindent

We present a model of quantum teleportation protocol based on  one-dimensional 
quantum dots  system. Three quantum dots with three electrons are used to perform
teleportation, the unknown 
qubit is encoded using one electron spin on quantum dot $A$, the other two dots $B$ and $C$ 
are coupled to 
form a mixed space-spin entangled state.  
By choosing the Hamiltonian for the mixed space-spin entangled system, we can
filter the space (spin) entanglement to obtain pure spin (space) entanglement
and after a Bell measurement, the unknown qubit is transfered to quantum dot $B$. 
Selecting an  appropriate Hamiltonian
for the quantum gate allows the spin-based information to be transformed into
a charge-based information. The possibility of generalizing this model 
to N-electrons is discussed.

\end{abstract}
\maketitle
\clearpage

Quantum teleportation is a technique for transferring quantum states from one place to another.  
A concise description of the protocol 
for teleporting a qubit can be  described as follows\cite{Benne}: 
The sender Alice has a source qubit which she wants to send to Bob 
and share with  Bob  an entangled pair,
an Einstein-Podolsky-Rosen (EPR) pair\cite{EPR}. Alice does a Bell measurement on 
the source qubit and her half of the EPR pair,
projecting the target qubit hold by Bob into 
a state  being  the same as the original state of the
source qubit up to a unitary transformation. Then Bob rotate the target qubit into 
the original state of the source qubit based on the two 
bits of classical information from Alice. 
The details of the protocol are shown in Fig.1: Alice performs a controlled-NOT (C-NOT) operation on 
her two qubits, using the source qubit as the 
control line. Then perform a Hadamard transformation on the source qubit.  
Alice then performs a measurement on her 
two qubits. After the measurement, 
Alice sends Bob two bits of classical information about the result of her  measurement, 
which is used by Bob to rotate the target qubit into the correct state. 
Quantum teleportation 
using pairs of entangled photons\cite{Bouw, Bos, Pan, Mar, Fat, Fur} and atoms\cite{Bar, Rie}  
have been demonstrated experimentally. There are also schemes  suggesting  using 
electrons to perform quantum teleportation\cite{Sau, Pas, Bee}.

In this paper, we study the quantum entanglement in an array of
quantum dots and  
propose a scheme to perform quantum teleportation in this system. We show that 
the space entanglement and spin entanglement contained in the quantum dots system modeled 
by the one-dimensional Hubbard Hamiltonian  
can be applied selectively to 
perform quantum teleportation.

Quantum entanglement is one of the most important concepts in quantum information theory and
quantum computation. It is 
key to the implementation of quantum information processing technology.
It has been realized that quantum entanglement can be used
as a controllable physical resource\cite{Nielsen}.
Theoretically, finding a measure for the 
 quantum entanglement is an important issue. For the fermion system, 
we choose Zanardi's measure\cite{Zanardi}, 
which is given in Fock
space as the von Neumann entropy.

Quantum dots system is one of the proposals for building a quantum computer\cite{Loss,Friesen}.
To describe the quantum dots, a simple approximation is to
regard each dot as having one valence orbital, the electron occupation could be
$|0>$, $|\uparrow>$, $|\downarrow>$ and
$|\uparrow\downarrow>$, with other electrons treated as core electrons\cite{Remacle}.
The valence electron can tunnel from a given dot to its nearest neighbor obeying the Pauli
principle and thereby two dots can be coupled together, this is the electron hopping effect. 
Another effect needs to be considered
is the on-site electron-electron repulsion.
A theoretical description
of an array of quantum dots can be modeled by the one-dimensional
Hubbard Hamiltonian:

\begin{equation}\label{initialHamiltonian}
{H} = -{t}\sum_{<ij>,\sigma}\ c_{i\sigma}^+\ c_{j\sigma} +
{U}\sum_{i} \ n_{i\uparrow} \ n_{i\downarrow}
\end{equation}

\noindent where $t$ stands for the electron hopping parameter, 
$U$ is the Coulomb repulsion parameter for electrons on the same site,
$i$ and $j$ are the neighboring site numbers, 
$\ c_{i\sigma}^+$ and $\ c_{j\sigma}$ are the creation and annihilation operators.

Entanglement using Zanardi's measure can be formulated as the von Neumann entropy given by

\begin{equation}
\label{initialHamiltonian}
{E_j} = -{Tr}(\rho_j log_2 \rho_j)
\end{equation}

\begin{equation}
\label{initialHamiltonian}
\rho_j = {Tr_j}(|\Psi> <\Psi|)
\end{equation}

\noindent where $Tr_j$ denotes the trace over all but the $j$th site and $|\Psi>$ is the antisymmetric
wave function of the fermion system. Hence $E_j$ actually describes the entanglement of the
$j$th site with the remaining sites.

In the Hubbard model, the electron occupation of each site has four possibilities, there
are four possible local states at each site, 
$|\nu>_j$ = $|0>_j$, $|\uparrow>_j$, $|\downarrow>_j$, 
$|\uparrow\downarrow>_j$. 
The entanglement of the $j$th site with the other sites is given by\cite{Gu}

\begin{equation}
\label{initialHamiltonian}
{E_j} = -{z} {Log}_2 {z}-{u}^+ {Log}_2 {{u}^+} - {u}^- {Log}_2 {{u}^-} -{w} {Log}_2 {w}
\end{equation}
where,
\begin{equation}
\label{initialHamiltonian}
\rho_j = {z}|0><0|+{u}^+|\uparrow><\uparrow|+u^-|\downarrow><\downarrow|+{w}|\uparrow\downarrow><\uparrow\downarrow|
\end{equation}
\begin{equation}\label{initialHamiltonian}
{w} = <{n}_{j\uparrow}{n}_{j\downarrow}>={Tr}({n}_{j\uparrow}{n}_{j\downarrow}\rho_j)
\end{equation}
\begin{equation}\label{initialHamiltonian}
{u}^+ = <{n}_{j\uparrow}>-w,\qquad  {u}^- = <{n}_{j\downarrow}>-w
\end{equation}
\begin{equation}\label{initialHamiltonian}
{z} = 1-{u}^+ - u^- - w=1-<{n}_{j\uparrow}>-<{n}_{j\downarrow}>+{w}
\end{equation}
The Hubbard Hamiltonian can be rescaled to have only one parameter $U/t$.

For the one-dimensional Hubbard model with half-filled electrons, we have 
 $<{n}_{\uparrow}>=<{n}_{\downarrow}>=\frac{1}{2}$, ${u}^+=u^-=\frac{1}{2}-w$
, and the local entanglement is given by

\begin{equation}\label{initialHamiltonian}
{E_j} = -2{w}{log}_2 {w}-2(\frac{1}{2}-{w}) {log}_2 {(\frac{1}{2}-{w})}
\end{equation}

For each site the entanglement is the same. 
Consider the particle-hole symmetry of
the model, we can see that $w(-U)=\frac{1}{2}-w(U)$, so the local entanglement is an even
function of $U$. As shown in Fig. 2, the minimum of the entanglement is $1$ as 
${U}\rightarrow\pm\infty$. As ${U}\rightarrow {+} \infty$
all the sites are singly occupied the only difference is the spin of the electrons on
each site, which can be referred as the spin entanglement. As ${U}\rightarrow {-} \infty$, 
all the sites are either
doubly occupied or empty, which is referred as the space entanglement. The maximum 
entanglement is $2$ at $U=0$, which is the sum of the spin and space
entanglement of the system. In  Fig. 2 we show the entanglement 
for two sites and two electrons, they
qualitatively agree with that of the Bethe ansatz solution for an array of sites\cite{Gu}.

Gittings and Fisher\cite{Gittings} showed that the entanglement in this system 
can be used in quantum teleportation. However,  in their scheme
both the charge and spin of the system are used to construct
the unitary transformation. 
In this paper, we propose a different  scheme to perform  quantum teleportation.
For  two 
half-filled coupled quantum dots, under the conservation of the total number of electrons $N=2$
and the total electron spin $S=0$, a quantum entanglement of $2$, two ebits 
can be produced according to 
Zanardi's measure. Let us describe the teleportation scheme using three cites, $A$, $B$ and $C$. 
Suppose the qubit $\alpha|\uparrow>+\beta|\downarrow>$ will be teleported from  site $A$, Alice,
to site $B$, Bob, where 
the two sites $B$ and  $C$ are in an entangled state,

\begin{equation}\label{initialHamiltonian}
|\Psi> = \frac{1}{\sqrt{2}}(\ c_{C\uparrow}^+{+}\ c_{B\uparrow}^+) 
\frac{1}{\sqrt{2}}(\ c_{C\downarrow}^+{+}\ c_{B\downarrow}^+)|0>.
\end{equation}

A spin-up electron and a spin-down electron are in a delocalized state on sites $C$ and
$B$. 
In the occupation
number basis $|\ n_{C\uparrow}\ n_{C\downarrow}\ n_{B\uparrow}\ n_{B\downarrow}>$,
the state of the system can be written as:

\begin{equation}
|\Psi> = \frac{1}{\sqrt{2}}(\ c_{C\uparrow}^+{+}\ c_{B\uparrow}^+) 
\frac{1}{\sqrt{2}}(\ c_{C\downarrow}^+{+}\ c_{B\downarrow}^+)|0>=
\frac{1}{2}(|0011>+|1100>+|1001>+|0110>). 
\end{equation}

From the state described by Eq. (10) we can see that in 
the basis of $|\ n_{C\uparrow}\ n_{C\downarrow}>$, 
there are four possible states: $|00>$, $|11>$, $|10>$, $|01>$.
Corresponding to each of the states on site $C$,  the states on site $B$ 
 are: $|11>$, $|00>$, $|01>$, $|10>$ in the
occupation number basis $|\ n_{B\uparrow}\ n_{B\downarrow}>$.
Under the restriction of the conservation of total number of electrons and total spin of the system, 
two ebits can be obtained, 
one is in the spatial degree of freedom, and the other is in the spin
degree of freedom. In the basis of
$|\ n_{C\uparrow}\ n_{C\downarrow}\ n_{B\uparrow}\ n_{B\downarrow}>$, the two
ebits are:

\begin{equation}\label{initialHamiltonian}
\beta_0 = \frac{1}{\sqrt{2}}(|1100>+|0011>),\qquad \beta_1 = \frac{1}{\sqrt{2}}(|1001>+|0110>)
\end{equation}

These two ebits can be used in quantum teleportation.
The C-NOT operation in the occupation number basis 
$|\ n_{A\uparrow}\ n_{A\downarrow}\ n_{C\uparrow}\ n_{C\downarrow}>$ is given by:

\begin{equation}\label{initialHamiltonian}
|1000> \leftrightarrow |1011>, |1010> \leftrightarrow |1001>, 
|01\ n_{C\uparrow}\ n_{C\downarrow}> \leftrightarrow |01\ n_{C\uparrow}\ n_{C\downarrow}>
\end{equation}

For the ebit $\beta_0$, in the
quantum teleportation process, in basis 
$|\ n_{A\uparrow}\ n_{A\downarrow}>|\ n_{C\uparrow}\ n_{C\downarrow}\ n_{B\uparrow}\ n_{B\downarrow}>$, 
as shown in Fig. 1, we have:

\begin{equation}\label{initialHamiltonian}
|\Psi_0> = (\alpha|10>+\beta|01>)\frac{1}{2}(|1100>+|0011>+|1001>+|0110>)
\end{equation}
\begin{equation}
|\Psi_1> = \alpha |10> \frac{1}{\sqrt{2}} (|0000>+|1111>)+
 \beta|01>\frac{1}{\sqrt{2}}(|1100>+|0011>)
\end{equation}
\begin{equation}
|\Psi_2> = \alpha(|10>+|01>)\frac{1}{2}(|0000>+|1111>)
 +\beta(|10>-|01>)\frac{1}{2}(|1100>+|0011>)
\end{equation}

When Alice does the measurement $M_1$ and $M_2$, the following results can be obtained:

\begin{eqnarray}
 & & |M_1M_2> \qquad \qquad |\ n_{B\uparrow}\ n_{B\downarrow}> {}
 \nonumber\\
 & & {}|1011> \qquad \qquad \alpha|11>+\beta|00> {}
 \nonumber\\
 & & {}|1000> \qquad \qquad \alpha|00>+\beta|11> {}
 \nonumber\\
 & & {}|0111> \qquad \qquad \alpha|11>-\beta|00> {}
 \nonumber\\
 & & {}|0100> \qquad \qquad \alpha|00>-\beta|11>
\end{eqnarray}

Then after doing a unitary transformation  using double electron occupation and zero electron 
occupation as basis, the source qubit can be obtain on site $B$. For this
system the Hamiltonian to perform the C-NOT operation is given by:

\begin{eqnarray}
H&=&|10>_A{}_A<10| (|11>_C{}_C<00|+|00>_C{}_C<11|)+ {}
 \nonumber\\
 & & |01>_A{}_A<01| (|11>_C{}_C<11|+|00>_C{}_C<00|) {}
 \nonumber\\
 & &=\frac{1}{2}(\sigma_{Z,A}+1)(\ c_{C\uparrow}^+\ c_{C\downarrow}^+{+}
 \ c_{C\uparrow}\ c_{C\downarrow}){+} {}
 \nonumber\\
 & &\frac{1}{2}(1-\sigma_{Z,A})(\ c_{C\uparrow}^+\ c_{C\downarrow}^+\ c_{C\uparrow}\ c_{C\downarrow}{+}
\ c_{C\uparrow}\ c_{C\downarrow}\ c_{C\uparrow}^+\ c_{C\downarrow}^+), 
\end{eqnarray}
where $\sigma_{Z,A}$ is the Pauli matrix. We can see that by using this Hamiltonian, the 
spin entanglement of the system is filtered, the space entanglement is used in the process.
An important point  is that the original state we try to teleport is in a superposition state
of electron spin
up and spin down. However, after the teleportation process, the state 
we obtain on site $B$ is a superposition
state of double electron occupation and zero electron occupation. 
The information based on spin has been 
transformed to information based on charge, but the information content is not changed. It is 
well known that a difficult 
task in quantum information processing and spintronics is 
the measurement of a single electron spin\cite{Radu}, 
in the scheme above, we changed the quantum information from spin-based to charge-based, thus 
makes the measurement fairly easier. This is also important in quantum computation based on 
electron spin since the readout can be easily measured.

For another ebit $\beta_1$, in the quantum teleportation process, in basis 
$|\ n_{A\uparrow}\ n_{A\downarrow}>|\ n_{C\uparrow}\ n_{C\downarrow}\ n_{B\uparrow}\ n_{B\downarrow}>$
, we have:

\begin{equation}\label{initialHamiltonian}
|\Psi_0> = (\alpha|10>+\beta|01>)\frac{1}{2}(1100>+|0011>+|1001>+|0110>)
\end{equation}
\begin{equation}
|\Psi_1> = \alpha |10> \frac{1}{\sqrt{2}} (|0101>+|1010>) +
 \beta|01>\frac{1}{\sqrt{2}}(|1001>+|0110>)
\end{equation}
\begin{equation}
|\Psi_2> = \alpha(|10>+|01>)\frac{1}{2}(|0101>+|1010>
 +\beta(|10>-|01>)\frac{1}{2}(|1001>+|0110>)
\end{equation}

When Alice does the measurement $M_1$ and $M_2$, the following results can be obtained:

\begin{eqnarray}
 & & |M_1M_2> \qquad \qquad |\ n_{B\uparrow}\ n_{B\downarrow}> {}
 \nonumber\\
 & & {}|1001> \qquad \qquad \alpha|01>+\beta|10> {}
 \nonumber\\
 & & {}|1010> \qquad \qquad \alpha|10>+\beta|01> {}
 \nonumber\\
 & & {}|0101> \qquad \qquad \alpha|01>-\beta|10> {}
 \nonumber\\
 & & {}|0110> \qquad \qquad \alpha|10>-\beta|01>
\end{eqnarray}

For this system the Hamiltonian to perform the C-NOT operation is:

\begin{eqnarray}
H&=&|10>_A{}_A<10| (|10>_C{}_C<01|+|01>_C{}_C<10|)+ {}
 \nonumber\\
& & {}|01>_A{}_A<01| (|01>_C{}_C<01|+|10>_C{}_C<10|) {}
 \nonumber\\
& &{}=\frac{1}{2}(\sigma_{Z,A}+1)(\ c_{C\uparrow}^+\ c_{C\downarrow}{+}\ c_{C\downarrow}^+\ c_{C\uparrow}) {}
 \nonumber\\
& & {}{+} \frac{1}{2}(1-\sigma_{Z,A})(\ c_{C\uparrow}^+\ c_{C\uparrow} \ c_{C\downarrow}
\ c_{C\downarrow}^+{+}\ c_{C\downarrow}^+\ c_{C\downarrow} \ c_{C\uparrow}\ c_{C\uparrow}^+)
\end{eqnarray}

Then after doing a unitary transformation  using the electron spin up and spin down as basis 
the source qubit can be recovered on site $B$. By using this Hamiltonian for the C-NOT operation
the space entanglement of the system is filtered, the spin entanglement is used in the process.
In the case of using $\beta_0$ or $\beta_1$ as ebits, the unitary transformation is performed in the
occupation number basis of
$|\ n_{B\uparrow}\ n_{B\downarrow}>$, using basis ${|11>, |00>}$ or 
${|10>, |01>}$, we can select the
basis separately, either charge or spin. We can also choose the Hamiltonians (one is
related to the spin entanglement and the other is related to space entanglement) for the 
C-NOT operation, when the Hamiltonian for one ebit is chosen, 
the ebit corresponding to the other Hamiltonian will be
filtered.

For $U \ne 0$, the state of
the $2$-electron $2$-sites system can be described as follows:

\begin{equation}\label{initialHamiltonian}
|\Psi> = a_{1}|1100> + a_{2}|0011> + b_{1}|1001> + b_{2}|0110>;  \;\; a_{1}^{2}+a_{2}^{2}+b_{1}^{2}+b_{2}^{2}=1,
\end{equation}

where $a_1=a_2$, $b_1=b_2$ because of the symmetry in the entangled pairs, such that the state
can be written as:

\begin{equation}
|\Psi> = a\beta_0 + b\beta_1; \;\; a^2+b^2=1.
\end{equation}

If $U>0$, the contribution of the spin entanglement to the total entanglement is greater
than that of the
space entanglement. The probability of getting the ebit $|\beta_1>$
 increases as $U$ becomes larger.
If $U<0$, the contribution of the space entanglement to the total entanglement becomes greater
than that of the spin entanglement, the probability of getting the ebit $|\beta_0>$ increases as $U$
becomes more negative. In the limit of $U$ goes to $ \pm \infty$, only spin entanglement or 
space entanglement will exist. This might be  related to the 
spin charge separation in the Hubbard model \cite{essler}.
In a previous study\cite{wang11}, we showed that the maximum
entanglement can be reached at $U>0$ by introducing asymmetric electron hopping impurity to the
system.
This is very convenient in the quantum information processing.
We can control the parameter $U/t$ to increase the probability of getting either ebit.

Here, we discussed implementing quantum teleportation in three-electron system. For 
more electrons and in the limit of ${U}\rightarrow {+} \infty$ there is no double occupation, the 
system reduced to the Heisenberge model, in the magnetic field.  The neighboring spins will
favor the  anti-parallel configuration for the ground state. If the spin at one end is flipped, then the spins 
on the whole chain will be flipped accordingly due to the spin-spin correlation. 
Such that the 
spins at the two ends of the chain are entangled, a spin entanglement, this can 
be used  for  
quantum teleportation, the information can be transfered through the chain. 
For ${U}\ne {+} \infty$, for the $N$-sites $N$-electron system with $S=0$, the first $N-1$ sites 
entangled with the $N$-th site in the same way as that of the two-electron two-sites system: if the 
$N$-th site has $2$ electrons, then the first $N-1$ sites will have $N-2$ electrons; if the
$N$-th site has $0$ electrons, then the first $N-1$ sites will have $N$ electrons; if the
$N$-th site has $1$ spin-up electron, then the total spin of  the first $N-1$ sites will be   
 $1$ spin-down; if the 
$N$-th site has the $1$ spin-down electron, then the total spin of  the first $N-1$ sites will
be {1} spin-up. So the same procedure discussed above  can be used  for quantum teleportation
but the new system with $N$-electrons is much more complicated than the previous
three electron system. Moreover, 
Alice needs to control the first $N-1$ sites and the source qubit. This situation is different
from the spin chain, this correlation can not be transfered from one end to the other.

In summary, we have studied the entanglement of an array of quantum dots modeled by the 
one-dimensional Hubbard Hamiltonian and its application in quantum teleportation. 
The entanglement in this 
system is a mixture of space and spin entanglement. The application of 
such entanglement in quantum teleportation process has been discussed, by applying different
Hamiltonians for the C-NOT operation, we can separate the ebit based on space entanglement
or spin entanglement and apply it in quantum teleportation process. It turns out that if
we use the ebit of the space entanglement, we can transform the spin-based quantum information
to the charge-based quantum information making the measurement fairly easy.

\begin{acknowledgments}
This work has been supported in part by the Purdue Research Foundation.
S.K. would like to acknowledge the financial support of the 
John Simon Guggenheim Memorial Foundation.
\end{acknowledgments}

\newpage

\begin{figure}
\begin{center}
\includegraphics[width=0.9\textwidth,height=0.3\textheight]{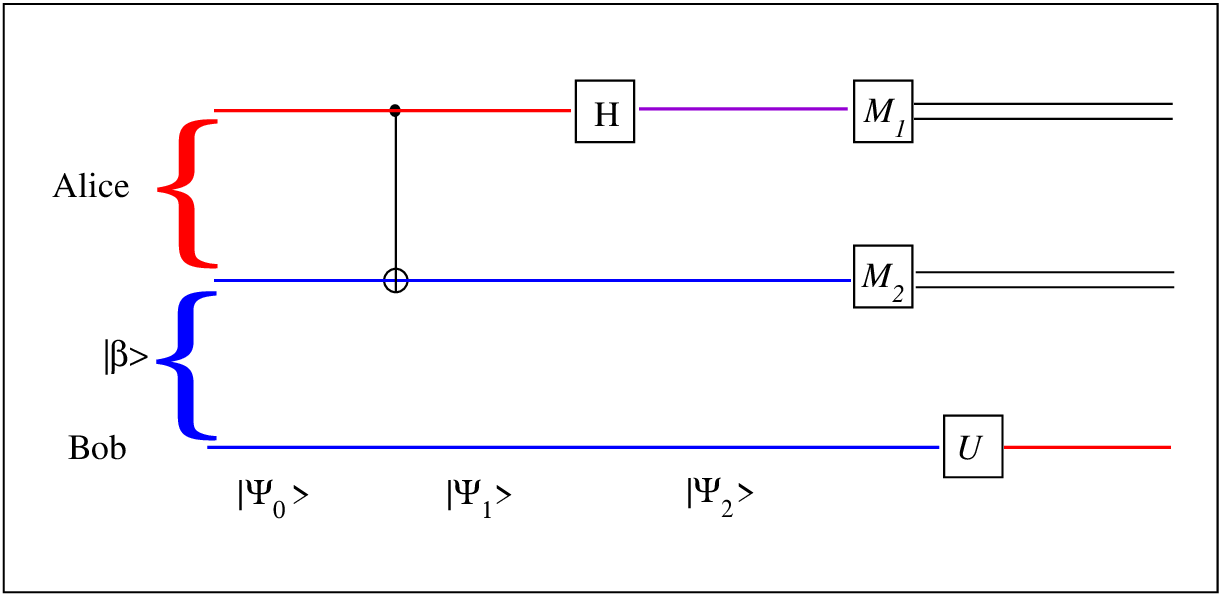}
\end{center}
\caption{Quantum circuit for teleporting a qubit. The two top lines represent Alice's system,
 while the bottom line is Bob's system. $H$ represents a Hadmard transformation, $M_1$ and
 $M_2$ represent the measurement on the two top lines.}
 \end{figure}

\begin{figure}
\begin{center}
\includegraphics[width=1.0\textwidth,height=0.5\textheight]{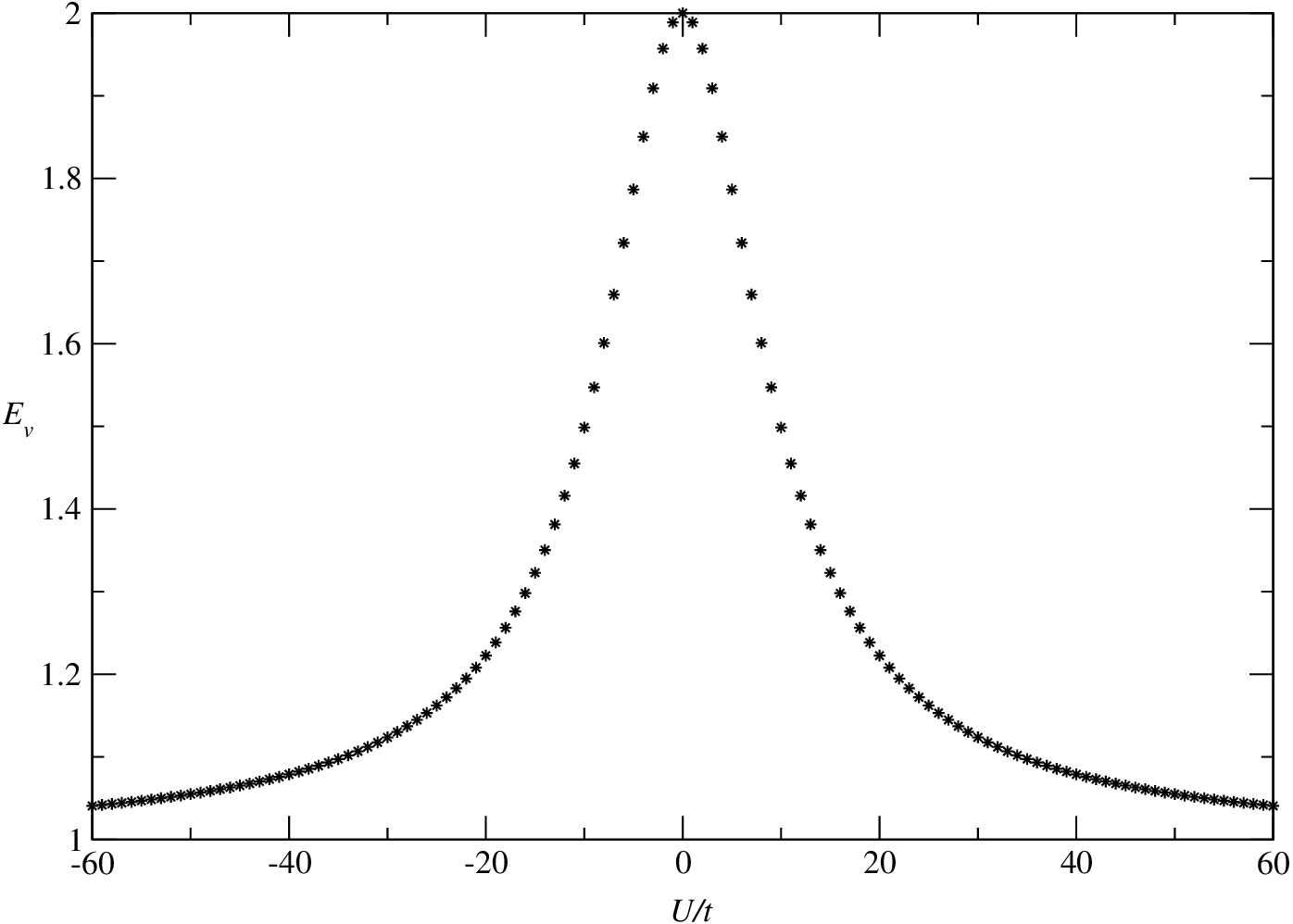}
\end{center}
\caption{ Local entanglement given by the von Neumann entropy $E_v$ versus $U/t$.}
 \end{figure}

\end{document}